\documentclass{emulateapj}

\usepackage{graphicx}
\usepackage{epstopdf}
\usepackage{epsfig}
\slugcomment {Accepted for publication in ApJ}

\shorttitle{DASCH Peculiar Nova-like Outburst}
\shortauthors{Tang et al.}

\begin{document}

%% LaTeX will automatically break titles if they run longer than
%% one line. However, you may use \\ to force a line break if
%% you desire.

\title{DASCH Discovery of A Possible Nova-like Outburst in A Peculiar Symbiotic Binary}

%% Use \author, \affil, and the \and command to format
%% author and affiliation information.
%% Note that \email has replaced the old \authoremail command
%% from AASTeX v4.0. You can use \email to mark an email address
%% anywhere in the paper, not just in the front matter.
%% As in the title, use \\ to force line breaks.

% JAO  Changed my name

\author{Sumin Tang\altaffilmark{1}, Jonathan E. Grindlay\altaffilmark{1},  Maxwell Moe\altaffilmark{1},
Jerome A. Orosz\altaffilmark{2}, \\
 Robert L. Kurucz\altaffilmark{1}, Samuel N. Quinn\altaffilmark{1}, Mathieu Servillat\altaffilmark{1}}
\altaffiltext{1}{Harvard-Smithsonian Center for Astrophysics, 60
Garden St, Cambridge, MA 02138}
\altaffiltext{2}{Astronomy Department, San Diego State University, 5500 Campanile Drive, San Diego, CA 92182-1221}
\email{stang@cfa.harvard.edu}

%% Mark off your abstract in the ``abstract'' environment. In the manuscript
%% style, abstract will output a Received/Accepted line after the
%% title and affiliation information. No date will appear since the author
%% does not have this information. The dates will be filled in by the
%% editorial office after submission.

%%JAO rearranged the line breaks

%%JAO changed the first sentence

\begin{abstract}
We present photometric and spectroscopic observations of a
peculiar variable (designated DASCH J075731.1+201735 or J0757) discovered
from our DASCH project using the digitized Harvard College Observatory
archival photographic plates.
It brightened by about 1.5 magnitudes in B within a year starting in
$1942$, and then slowly faded back to its pre-outburst brightness
from 1943 to the 1950s.
The mean brightness level
was stable before and after the outburst, and ellipsoidal variations with
a period of $P=119.18\pm0.07$ days are seen,
suggesting that the star is tidally distorted.  Radial-velocity measurements
indicate that the orbit is nearly circular ($e=0.02\pm 0.01$) with
a spectroscopic
period that is the same as the  photometric period.
The binary consists of a $1.1\pm0.3\ M_\odot$ M0III star,
and a $0.6\pm0.2\ M_\odot$ companion, very likely a white dwarf (WD).
Unlike other symbiotic binaries, there is no sign of
emission lines or a stellar wind in the spectra.
With an outburst timescale of
$\sim10$ years and estimated B band peak luminosity $M_B\sim0.7$,
J0757 is different from any other known classic or symbiotic novae.
The most probable explanation of the outburst is Hydrogen shell-burning
on the WD, although an accretion-powered flare cannot be ruled out.
\end{abstract}

%% Keywords should appear after the \end{abstract} command. The uncommented
%% example has been keyed in ApJ style. See the instructions to authors
%% for the journal to which you are submitting your paper to determine
%% what keyword punctuation is appropriate.


%% From the front matter, we move on to the body of the paper.
%% In the first two sections, notice the use of the natbib \citep
%% and \citet commands to identify citations.  The citations are
%% tied to the reference list via symbolic KEYs. The KEY corresponds
%% to the KEY in the \bibitem in the reference list below. We have
%% chosen the first three characters of the first author's name plus
%% the last two numeral of the year of publication as our KEY for
%% each reference.

\section{Introduction}
Symbiotic stars are interacting binaries consisting of an evolved red giant and
a hot object, which is usually a White Dwarf (WD; see e.g. Kenyon 1986, 1994; Nussbaumer 2000; Miko{\l}ajewska 2007 and references therein).
Mass is transferred from the red giant to its hot companion via stellar wind in most cases,
but could also be Roche lobe overflow in some cases.
They usually show low temperature absorption features from the red giant,
and strong emission lines from surrounding circumstellar material ionized by the hot component.
Most symbiotic stars have red giant masses $0.6-3.2\ M_\odot$,
hot component masses $0.4-0.8\ M_\odot$, and orbital periods $\sim200-1000$ days (Miko{\l}ajewska 2003).

Symbiotic stars are highly variable sources with variation timescales
from minutes to decades.  The most common type of outburst, called
a classical symbiotic outburst, recurs multiple times on decades-long
time scales with
amplitudes of $1-3$ mag, and with each outburst lasts from months to a few
years (Kenyon 1986; Miko{\l}ajewska 2011).
The nature of these
outbursts is not entirely clear, and may be related to instabilities
in the accretion disks (Miko{\l}ajewska 2003; Sokoloski et al. 2006).  Another relatively
rare type of outburst is symbiotic nova (also called slow nova), which
shows a single long outburst lasting from years to decades due to
thermonuclear runaways at the surface of the WD (Allen 1980;
Paczy\'{n}ski \& Rudak 1980; Kenyon \& Webbink 1984).  The peak
luminosities range from 4000 to 40,000 $L_\odot$, depending on the
mass of the WD (M\"{u}rset \& Nussbaumer 1994; Iben 2003).  They mimic
classical novae but with much longer timescales.  Only 9 symbiotic
novae are known so far (Belczy\'{n}ski et al. 2000; Miko{\l}ajewska
2011).

DASCH (Digital Access to a Sky Century @ Harvard)
is a project to
digitize and analyze the scientific data contained in the Harvard
College Observatory (HCO) photographic
plates taken from the 1880s through the 1980s.
The motivation is to explore the temporal
variations of stars and active galaxies on the relatively poorly explored
$\sim1-100$ yr timescales (but extending down to $\sim1$ week timescales as well),
as summarized by Grindlay et al.\
(2009 and 2012, in preparation). We developed the astrometry and photometry pipeline (Laycock et al.\ 2010;
Los et al.\ 2011; Servillat et al.\ 2011; Tang et al. in prep.),
and have scanned $\sim$ 19,000 plates in five initial fields.  Here we report the discovery
of an unusual 10-yr outburst in a peculiar symbiotic system, DASCH
J075731.1+201735 (named after its equatorial J2000
coordinate; hereafter J0757).

\section{Observations and Results}

\subsection{DASCH and ASAS light curves}
J0757 was measured on $694$ plates near the M44 field
noted by its
peculiar long-term variability.  These plates cover 5$-$25 degrees on
a side with typical limiting magnitudes $B\sim14-15$ mag, and most of them
are blue sensitive emulsions.
We used the GSC2.3.2 catalog (Lasker et al. 1990) for
photometric calibration, and our typical uncertainty is $\sim0.1-0.15$
mag (Laycock et al. 2010; Tang et al. in preparation).  There are
$\sim1.2\times10^5$ objects with more than 100 magnitude measurements
distributed over $\sim 100$ years from these plates covering the M44 field.  J0757, which is
N2211021132 in the GSC catalog, is the only one found with $>1$
magnitude outburst on a $\sim 10$ year timescale above a well-defined
quiescence level.

The DASCH light curve of J0757 is shown in Figure 1.  We supplement the
figure with V-band data from
ASAS starting in the year 2000 (J0757 has
the designation ASAS J075731+2017.6; Pojmanski 2002).
J0757 is classified by ASAS as
a semi-detached/contact binary with best-fit period of 119.2 days
with 0.16 mag variations in the V band.  Since ASAS data are in the V
band, while DASCH magnitudes are in B, we added 1.5 mag to the ASAS V mag
in the plot, which is about the sum of the maximum galactic extinction
($E(B-V)=0.06$; Schlegel et al. 1998) and the typical $B-V$ value for
a M0III star (1.43; Pickles 1998).
We will discuss the spectral
classification in section 2.2.

As shown in Figure 1, J0757 was in quiescence until about March 1942.
J0757 then brightened over the rest of the year of 1942,
and slowly decayed until about 1950.
The peak magnitude was reached around January, 1943.
There is a seasonal gap between April and October in 1942 when the star
was low or not visible in the night sky, and thus we do not know when
exactly the outburst started.  If we naively assume that its magnitude
increased linearly,
then the outburst rise time is $\sim 88.5$
days (see the lower right panel in Figure 1).

%
%   JAO simplified the discussion here
%
%and passed through the mean of the two points at
%JD 2430680 and 2430701 in a intermediate bright state, and the first point
%in the brightest state (JD 2440731), as shown in blue dashed line in
%the lower-right panel in Figure 1, .

%
%  JAO redid the line breaks
%

%
%  JAO changed below
%
Before and after the outburst when data are available
(i.e.\ 1890$-$1942 and $1962-1990$), the magnitudes of J0757
were reasonably stable (note that the relatively constant B$\sim11.9$ from $\sim1950-1954$
 was not yet the quiescent value). The median DASCH magnitude before the outburst (e.g.\ before June
1942) is 12.24 and  the median magnitude
after the outburst (after 1960) is 12.26.  The median magnitude
of the entire DASCH light curve excluding the period between June,
1942 and the start of 1960 is 12.25, with rms scatter = 0.14 mag for the 607 data points outside the outburst.
The median ASAS V magnitude is 10.75 mag, which
corresponds to B=12.24, assuming the
typical colors of an M0III star with galactic extinction.

\subsection{Light curve folding and the photometric period}
Both the DASCH and ASAS light curves show ellipsoidal variations.
The ellipsoidal variation amplitude is $\sim0.11$ mag in the DASCH data (B band, 1890$-$1942 and $1962-1990$),
and $\sim0.16$ mag in the ASAS data (V band, $2003-2009$).

We folded the light curves of J0757 to search for its photometric period.
We divided DASCH light curve into two phases,
i.e. before the outburst and after the outburst, as described in section 2.1.
We then got four sets of light curve data: ASAS light curve,
DASCH light curve excluding the outburst (before or after the outburst), DASCH light curve before the outburst, and DASCH light curve after the outburst.

We first folded the ASAS light curve with a given period, and fit it with a 3 harmonics Fourier model (an example is shown as the blue line in the upper panel of Figure 2).
We fixed the model line in shape, but allow two free parameters, i.e. median magnitude and amplitude, for adjustments.
Next, we fit the four sets of light curve data to the 3 harmonics Fourier model,
and derived the reduced $\chi^2$ as a function of different trial periods, as shown in the lower panels in Figure 2.
DASCH data were binned by phase at the given period in the fitting.
Our best fit period is $119.18\pm0.07$ days.
The fact that the binary period is unchanged (within the errors) sets limits on the change of the orbit
(due to mass transfer or onset of common envelope evolution).
The resulting value for $\dot{a}/a$ is $0\pm4\times10^{-6}$ yr$^{-1}$.

%
%  JAO filled in the line breaks below
%

\subsection{Optical Spectroscopy}
Spectra were obtained with the FAST spectrograph (Fabricant et
al. 1998) on the 1.5-m Tillinghast reflector telescope at the
F. L. Whipple Observatory (FLWO) for spectral classification.  The
spectra were reduced and
wavelength-calibrated with standard packages (Tokarz \& Roll
1997).  Figure 3 shows the FAST 300 grating spectrum,
which has a resolution of about 7~\AA.
The upper panel
shows the whole spectrum, which suggests that its spectral type is M0;
The lower panels show two luminosity indicators, which suggest that
DASCH J0757 belongs to luminosity class III.  The lower-left panel
shows Ca I at 4226 \AA, which disappears with increasing luminosity,
indicating that DASCH J0757 is not a luminosity class I star.  The
lower-right panel shows the CaH B-band at 6385 \AA, and the blend near
6362 \AA
\ consisting of Ti, Fe, and Cr.  The weakness of CaH B-band
suggests that DASCH J0757 is not a main sequence star.
We therefore adopt a spectral type of
$\sim$M0III for DASCH J0757,
with uncertainties of $\sim \pm1$ in both spectral type and luminosity class.

%
%  JAO reworded slightly above
%

%
%   JAO filled  in the line breaks below
%

We also obtained spectra using the MIKE echelle spectrograph
(Bernstein et al. 2003) on the 6.5 m Magallen/Clay telescope with a $0.7$'' $\times5$''
slit on January 17, 2011.  The spectral resolution is $R\sim30,000$.  The
spectra were extracted using an IDL pipeline developed by S. Burles,
R.  Bernstein, and J. S.
Prochaska\footnote{\texttt{http://web.mit.edu/$\scriptstyle\sim$burles/www/MIKE/}}.  The
reduced spectra have $S/N\sim50-100$ in 4300-9000 \AA\ and
$S/N>\sim10$ in 3700-4300 \AA.
% BK wrote the following
We computed a grid of spectrum templates for a range of abundances,
effective temperatures, and gravities using Kurucz's programs ATLAS12
and SYNTHE (Kurucz 2005).  We derived the projected rotation velocity
of $10\pm1$ km s$^{-1}$.  In M giants the continuum opacity is mainly H-
which varies with metal abundance.  The ratio of line to continuum does
not change strongly with abundance. The TiO background is strong at any
abundance because both Ti and O are alpha-enhanced at low abundances.
We estimated an effective temperature of $T_{eff}=3850\pm50$ K; a surface
gravity of $log\ g=1.0\pm0.5$ cgs; and a metallicity of $[Fe/H]=-0.6\pm0.1$,
assuming J0757 is a population II star with alpha enhancements.
The spectra can also be fit with Population I solar abundance templates.
There is no sign of emission lines or wind in its spectra.

%
%   JAO changed above
%

To measure the radial velocity curve
of J0757, we obtained spectra on 15 different
nights from October, 2010 through April,
2011 with the Tillinghast Reflector
Echelle Spectrograph (TRES; Szentgyorgyi \& Fur\'{e}sz 2007) on FLWO
1.5 m.  These spectra were reduced and
wavelength-calibrated with standard
packages (Mink 2011).  The spectral resolution is $R\sim44,000$.  The
reduced spectra have typical S/N $\sim10-30$ in 4500-9000 \AA, and S/N
$\sim5-10$ in 4000-4500 \AA.

Radial velocities of DASCH J0757 were derived from the TRES spectra, and
were fitted to a model velocity curve
using the derived photometric period (see Figure 4).
%
%  JAO These belong in the figure caption
%
%as a function of orbital phase, as shown in Figure 4.
%Blue solid dots are observation data using the TRES spectrograph at
%the FLWO 1.5m, and the smooth curve is the best-fitting model.
%Residuals from the fit are given in the lower panel.
The derived orbital
parameters are listed in Table 1.
The resulting mass function is
\begin{equation}
f_1(M_2)=(3.31\pm0.02)\times10^{-2}\ M_\odot ,
\end{equation}
and
\begin{equation}
M_2 \sin i = (0.3211 \pm 0.0006) (M_1+M_2)^{2/3}\ M_\odot ,
\end{equation}
where subscript $1$ denotes the red giant, and $2$ denotes its companion.

For comparison, we have also fit a model velocity curve without using the photometric
period
of $P=119.18\pm0.07$ days, and got similar results:
$T_0 = 2455597.969\pm0.06$,
$P   = 118.89\pm0.080 $ days,
$\gamma   = -33.57 \pm 0.04$  km s$^{-1}$,
$K   = 13.885\pm 0.023$ km s$^{-1}$,
$e   = 0.024\pm 0.015$,
and $\omega   = 70.38 \pm 0.09$ deg.
In the following discussion throughout the paper, we adopt the
parameters obtained using the
photometric period of $P=119.18\pm0.07$ days, as listed in
Table 1.

%
%  JAO Move this to section 6
%
%\subsection{Ellipsoidal variation}
%As shown in Fig. 5.
%
%

\subsection{SED from archival data}

Using archival data from WISE, IRAS, AKARI/IRAC, 2MASS, FON, SDSS,
GALEX and ROSAT, we plotted the spectral energy distribution (SED) of
J0757, as shown in Figure 5.  The phase of the GALEX observation is
$0.603\pm0.01$, and the phase of ROSAT observation is $0.426\pm0.02$.
Here phase 0.5 is defined as inferior conjunction of the red giant.
The SED is well
matched by a synthetic spectrum of a $T_{\rm eff}=3750$ K, and $\log g =
1.5$ giant (Lejeune et al. 1997), from IR to optical, as shown in
black line in Figure 6.  However, there is an excess in the GALEX NUV
($\sim1900-2700$\AA), which suggests a hot component in addition to
the giant.

%
%   JAO change the section to ``system parameters''
%

\section{System parameters}

\subsection{Ellipsoidal models}

We derive system parameters from the light
and velocity curves.
For this purpose
we used the ELC code of Orosz \& Hauschildt (2000) to model
the ASAS V-band light curve and the TRES radial velocity curve.
The ELC model is based on standard Roche geometry
and uses specific intensities from model atmospheres, in particular
the {\sc NextGen} models of Hauschildt et al.\ (1999).   ELC's
use of the model atmosphere specific intensities eliminates the need
for a parameterized limb darkening law, which is important in
cases with red giants where the limb darkening near the stellar
limb is very nonlinear (Orosz \& Hauschildt 2000).   We don't have
model atmosphere specific intensities appropriate for white dwarfs.
However, since the red giant dominates the V-band flux, this is not
a serious limitation.

In our initial runs, we had 8 free parameters:  the orbital period
$P$, the time of periastron passage $T_0$, the inclination $i$,
the Roche lobe filling factor $f_1$, the eccentricity $e$, the
argument of periastron $\omega$, the $K$-velocity of the
red giant $K$, and the mass of the red giant $M_1$.  We assumed
synchronous rotation.  The models are somewhat insensitive to
the temperature of the red giant, and that parameter was held fixed
at $T_1=3720$ K.  In addition to the light and velocity curve, we used
the projected rotational velocity of the red giant as an extra
constraint.

The model fits were optimized using ELC's genetic code and
its Monte Carlo Markov Chain code.  After inspection of the
post-fit residuals, additional free parameters were considered.
We allowed the red giant's rotation to be asynchronous with the
orbit, where $\Omega$ is the ratio of the spin frequency to the
orbital frequency.  We also
accounted for heating by the WD.  In this case, we fixed the
luminosity of the irradiating source to $\log L_i=36.6$
(c.g.s. units) and allowed the albedo $A_1$ to be a free parameter
(the choice of $A_1$ is somewhat arbitrary since
different combinations of $L_i$ and $A_1$ can give identical light curves).
Finally, we included an accretion disk that is parameterized
by the outer radius in terms of the Roche radius $r_{\rm out}$,
the temperature at the inner rim $T_d$, and the exponent of the
disk temperature distribution $\xi$, where $T(r)=T_d(r/r_i)^{\xi}$.
%For models using the accretion disk, we used black bodies and
%a standard limb darkening law since our table of model atmosphere
%specific intensities does not have the needed range of temperatures.

Some typical results are shown in Figure 6.  The binned
ASAS light curve has maxima with different heights,
which could be indicative of an eccentric orbit.  However, the
simultaneous fit to the radial velocities indicates a
small eccentricity (0.012).  Also, the minimum near phase
0.5, corresponding to the inferior conjunction of the red giant,
is deeper than the minimum at phase 0.0.  For a pure ellipsoidal
light curve, the minimum at phase 0.0 would be the deeper one.
Heating by the irradiating source ``fills in'' the minimum near phase
0.   However, in all of the models considered, we never had a case
where the model curve had a deeper minimum near phase 0.5, so simple
heating cannot explain the difference in the observed minima.

An accretion disk could potentially eclipse the red giant near phase
0, thereby making the minimum there deeper.  This is opposite of
what we want.  Normally, the accretion disk does not give phase
modulated light in the ELC model (if it is not eclipsed by the star).
However, a ``hot spot'' on the disk rim can be used to  mimic
the place where the matter transfer stream impacts the disk.  As the
binary turns in space, different areas of this spot are visible, giving
rise to extra light at certain phases.   An example model is
shown in Figure 6.  The hot spot has an angular width of $18^{\circ}$,
an azimuth (measured from the line of centers in the direction or the
orbital motion) of $66^{\circ}$, and has a temperature of
3800 K, which is 4.3 times hotter than the
rest of the disk rim.
The maximum near phase 0.75 is fit, but
the minimum near phase 0.5 is still not well-fit.

We did two runs of fitting with a Monte Carlo Markov chain,
one with an external constraint on the rotational velocity,
and one without the constraint.
Results from both runs are similar.
The best-fit ellipsoidal models have inclination of $60^{\circ}\pm5^{\circ}$,
WD mass of $0.5\pm 0.1\, M_{\odot}$,
red giant mass of $1\pm 0.3\,M_{\odot}$,
and red giant radius of $43\pm 4\,R_{\odot}$.

We note that in order to match the observed rotational velocity of $V_{rot} sini=10\pm1$ km
s$^{-1}$ derived from the MIKE spectra, the red giant must
be rotating 0.7 times slower than its synchronous velocity.
Given the typical tidal synchronization timescale of $\sim10^5$ yr (Zahn 1989) adopting the
parameters listed in Table 1,
the red giant is expected to be co-rotating with the binary.
Consequently, the size of the giant
should be $R_1=PV_{rot}/2\pi=23.5/\sin i\ R_{\odot}$, which is $27\ R_{\odot}$ if
$i=60^{\circ}$, which are smaller than the radius derived from ellipsoidal fitting.
Such a discrepancy has been seen in many symbiotic binaries,
and is an unresolved ``embarrassing problem'' (Miko{\l}ajewska 2007).
It probably comes from the extended atmosphere,
and that limb darkening near the limb is very nonlinear.
The light curve models with near-linear limb darkening produce ellipsoidal light curves with amplitudes that are too small.
Also, the rotational broadening kernel can be a lot different than the usual analytic kernel,
which in turn leads to strange biases in the $V_{rot} sini$ value (Orosz \& Hauschildt 2000).

\subsection{Estimates from the spectroscopic observations}

Here we estimate the astrophysical parameters of DASCH J0757,
based on the results in Sections 2 and 3.1.
Since the galactic extinction is small ($E(B-V)=0.06$; Schlegel et al. 1998),
we do not take it into account for simplicity.
The following estimated parameters are summarized in Table 1.

The inclination from the ellipsoidal fitting is $i\sim 55^{\circ}-65^{\circ}$,
and we adopt a more tolerant value of $i=45^{\circ}-75^{\circ}$ in the following estimates.
The radius of the red giant ranges from $\sim25-33$ $R_{\odot}$ (from $V_{rot}$ assuming synchronized rotation)
to $\sim43$ $R_{\odot}$ (from ellipsoidal fitting), as discussed in Section 3.1.
Therefore, we adopt a radius and error bar that encompasses both methods as $R_1=35\pm9$ $R_{\odot}$.
At $T_{eff}=3850\pm50$ K, the bolometric luminosity of the red giant is $L=250\pm130$ $L_{\odot}$,
and its distance is $1.0\pm0.3$ kpc.
The absolute magnitude of the system in quiescence is then $M_B=2.2\pm0.6$.

Combining the radius, effective temperature, $log\ g=1.0\pm0.5$ cgs and $[Fe/H]=-0.6\pm0.1$,
the red giant mass inferred from stellar evolutionary models (Bertelli et al. 2008) is
$M_1\sim0.9\pm0.4 \ M_{\odot}$.
This mass estimate for the giant is determined solely from the equations of stellar structure and is therefore independent of its prior evolution.
The mass of the giant may have been greater in the past.
However, considering the giant is only partially filling its Roche lobe as inferred from the ellipsoidal fitting,
and given the M0III wind loss rate of $\sim10^{-9}-10^{-8}\ M_\odot$ yr$^{-1}$ (Cox 2000),
we do not expect it has lost a significant fraction of its current mass over the $\sim10^7$ yr lifetime of the giant in its current phase of evolution.
Requiring 0.8 $M_\odot$ to turn-off from main sequence within a Hubble time,
we then have $M_1\sim1.1\pm0.3 \ M_{\odot}$.
From equation (2) and our adopted  $i=45-75^{\circ}$, this implies the
secondary mass is $M_2\sim0.6\pm0.2\ M_{\odot}$, and the orbital separation
$a\sim120\pm11\ R_{\odot}$.  Therefore, the Roche lobe radii of the two
stars are $R_{RL,1}\sim53\pm4\ R_{\odot}$, and
$R_{RL,2}\sim38\pm6\ R_{\odot}$, for the M giant and the secondary,
respectively (Eggleton 1983).  The Roche lobe filling factor of the M
giant is then $f_{RL}=R_1/R_{RL,1}=0.66\pm0.17$.

If the deviation of the ASAS folded light curve and the ellipsoidal
model with a giant alone is caused by a hot component due to a WD companion, its accretion disk and/or a tidal stream,
then it is about
3\% the luminosity of the giant in the V band.
The luminosity of the hot component is then $2-8\ L_{\odot}$,
for an effective temperature in the range of $3800-10,000$ K.
For a $M_2\sim0.6\ M_{\odot}$ WD with $R_2\sim0.01$ $R_{\odot}$, assuming all the gravitational
potential energy is converted to thermal radiation, the corresponding
accretion rate is $1-4\times10^{-9} \ M_\odot$ yr$^{-1}$.

During the 1942-1950s outburst, the B magnitude of J0757 increased by
$\sim1.5$ mag at the peak.  At a distance of $D\sim1$ kpc, the
absolute B magnitude of the outburst is $M_{B, outburst\ peak}\sim0.7$
mag.  The peak bolometric luminosity of the outburst,
$L_{outburst\ peak}$ depends on its SED.  If the outburst has similar
SED as a M0III star, then $L_{outburst\ peak} \sim 750\ L_\odot$.
The minimum required bolometric luminosity would be when the radiation is
peaked in the B band, i.e. $L_{outburst\ peak} \sim 100\ L_\odot$ at
$T\sim8100$ K.
Such a temperature is consistent with the observed hot component temperature of CH Cyg during its 1992 outburst powered by accretion (Skopal et al. 1996).
If the outburst is powered by Hydrogen shell-burning,
the temperature during the outburst would be much hotter,
i.e. $T_{outburst}\sim10^5$ K as observed in some symbiotic stars (M\"{u}rset et al. 1991).
For example, Z And is found to have a temperature of $\sim9\times10^4$ K during its 2000-2002
outburst (Sokoloski et al. 2006).
  If the outburst in J0757 has a temperature
of $9\times10^4$ K, most of the radiation is in the UV, and the corresponding
bolometric luminosity of $\Delta B=1.5$ mag increase is $L_{outburst\ peak}
\sim 2.5\times10^4\ L_\odot$ assuming black body radiation.

\begin{table*}[tb]
\caption{Astrophysical parameters for DASCH J0757. Details are discussed in Sections 2\&3.  \label{tbl-1}} \centering \leavevmode
\begin{minipage}{\textwidth}
\tabcolsep 3.2pt \footnotesize
\begin{tabular}{ll}
\tableline Parameter  & Value \\
\tableline
Observational parameters with highest confidence: & \\
\tableline
Spectral type & M0III  \\
$T_0$ (HJD) & $2455597.980 \pm 0.054 $ \\
Period (days) & $119.18\pm0.07$ \\
$\gamma$ (km s$^{-1}$) & $-33.57\pm0.04$ \\
$K_{1}$ (km s$^{-1}$) & $13.906\pm0.025$ \\
Eccentricity & $0.025\pm0.01$ \\
$\omega$ (deg) & $70.49\pm0.09$ \\
$V_{rot} \sin i$ (km s$^{-1}$) & $10\pm1$ \\
\tableline
Parameters derived from atmosphere fitting: & \\
\tableline
$T_{eff}$ (K) & $3850\pm50$ \\
log\ g (cgs) & $1.0\pm0.5$ \\
\texttt{[Fe/H]} & $-0.6\pm0.1$  \\
\tableline
Estimates based on models: & \\
\tableline
$i$ (deg) & $60\pm15$ \\
$M_1$ ($M_{\odot}$) & $1.1\pm0.3$ \\
$M_2$ ($M_{\odot}$) & $0.6\pm0.2$ \\
Orbital seperation ($R_{\odot}$) & $120\pm11$ \\
$R_1$ ($R_{\odot}$) & $35\pm9$ \\
$L_1$ ($L_{\odot}$) & $250\pm130$ \\
Distance (kpc) & $1.0\pm0.3$ \\
$h_z$ (pc) & $400\pm110$ \\
$M_B$ in quiescence & $2.2\pm0.6$ \\
$M_B$ outburst & $0.7\pm0.6$ \\
$L_{\texttt{outburst peak}}$ ($L_{\odot}$) & $10^2-10^4$ (if no constraint on $T_{hot}$) \\
 & $100$ (if $T_{hot}\sim 8100$ K)\\
 & $2.5\times10^4$ (if $T_{hot}\sim 90,000$ K)\\
$L_{hot}$ in quiescence ($L_{\odot}$) & $\sim2-8$ \\
$\dot{M}$ in quiescence ($M_{\odot}$ yr$^{-1}$) & $\sim1-4\times10^{-9}$ \\
\tableline
\end{tabular}
\end{minipage}
%% Any table notes must follow the \end{tabular} command.
% \label{outcomes} \tablenotetext{1}{From its FAST spectra by comparing with standard stars.}
\end{table*}

\section{Discussion}

The amplitude and timescale of the 1940s outburst,
and its stability during quiescence phases (except the periodic ellipsoidal variation),
is very unusual.
The mass loss rate of a M0III star is $\sim10^{-9}-10^{-8}\ M_\odot$ yr$^{-1}$ (Cox 2000).
With a Roche lobe filling factor $\sim0.5-0.8$ as discussed in section 3.2,
a significant fraction of the wind could be transferred to its companion via wind Roche lobe overflow (Podsiadlowski \& Mohamed 2007).
If the secondary is a 0.6 $M_{\odot}$ Main Sequence (MS) star with $R\sim0.6$ $R_{\odot}$,
the only way to power the outburst is by gravitational energy release.
The least required accretion rate to power the 1940s outburst of J0757
is then $3\times10^{-6}\ M_\odot$ yr$^{-1}$ ($L_{peak}\sim100\ L_{\odot}$ if $T\sim8100$ K),
which is 2 to 3 orders of magnitude higher compared with the typical mass loss rate of the M giant.
Therefore, a compact object, almost certainly a WD (at $\sim0.6$ $M_{\odot}$, it cannot be a Neutron star or a black hole), is needed to power the outburst.

If the outburst was powered entirely by accretion on to a 0.6 $M_{\odot}$ WD,
the least required accretion rate to power the 1940s outburst of J0757
is $5\times10^{-8}\ M_\odot$ yr$^{-1}$, assuming all the potential energy
is converted into radiation ($L_{peak}\sim100\ L_{\odot}$ if $T\sim8100$ K).
However, J0757 only showed one single outburst and was stable before and after the outburst,
which is different from other symbiotic systems with accretion-powered outbursts,
such as PU Vul (before its nuclear runaway outburst in 1977),
which was highly variable with multi-epoch outbursts (Liller \& Liller 1979).
Another symbiotic variable powered by accretion, CH Cyg,
was stable prior to 1963 for several decades,
but later showed a series of outbursts since the onset of activity in 1963 (Miko{\l}ajewski et al. 1990).

The most probable explanation of the J0757 outburst, is then Hydrogen burning on the surface of the WD.
The outburst profile of J0757, i.e. a relatively sharp rise and exponential-like decay,
closely resembles that of Z And and CI Cyg,
which are believed to be related to the presence of unstable accretion disk around a white dwarf with hydrogen burning shell (Mikolajewska 2003, et al. 2002).
A ``combination nova'' outburst on Z And has been reported by Sokoloski et al. (2006),
which is triggered by an accretion instability and resulted in increased thermonuclear burning.
The outburst of J0757 may have started at the stable burning zone
($\dot{M}>\dot{M}_{crit}=3.3\times10^{-8}\ M_\odot$ yr$^{-1}$ for a $0.6\ M_\odot$ WD; Nomoto et al. 2007; Shen \& Bildsten 2007)
with an increased accretion rate due to an accretion disk instability (King et al. 2003),
when the accumulated envelope mass on the WD exceeded the critical value
($3\times10^{-5}\ M_\odot$; Townsley \& Bildsten 2004).
Then, consequently, hydrogen was ignited and started quasi-steady burning to power the outburst.
The luminosity of such a quasi-steady burning increases with higher WD mass, as well as higher accretion rate.
For a $0.6\ M_\odot$ WD, the theoretical luminosity at $\dot{M}_{crit}$ is $L\sim10^{3.4}\ L_\odot$,
and the effective temperature is $\sim2.5\times10^5$ K;
If $\dot{M}\sim4\dot{M}_{crit}$, then the luminosity is $\sim 10^{4}\ L_\odot$,
and the effective temperature is $\sim8\times10^{4}$ K (Nomoto et al. 2007).
Such a range of luminosity and corresponding temperature is consistent with the estimated peak luminosity of J0757,
which is $\sim10^4\ L_\odot$ if $T\sim10^5$ K.

The timescale of the Hydrogen shell burning strongly depends on the mass of the WD,
accretion rate, and whether a optically thick wind occurs (Kato \& Hachisu 1994; Prialnik \& Kovetz 1995).
Kato \& Hachisu (2011) suggested that the companion star plays an important role in producing a wind,
and thus affects the timescale of the outburst.
As shown in Figure 6 in Kato \& Hachisu (2011),
in wide symbiotic binaries like PU Vul ($P_{orb}=4900$ days),
the expanded WD envelope is not affected by the companion,
resulting in a quiet evolution without strong winds in the outburst,
and thus longer outburst (peak plateau timescale of $\sim10$ yr);
On the other hand, in close binaries such as V723 Cas and HR Del ($P_{orb}<1$ day) with similar WD masses as PU Vul ($\sim0.6\ M_{\odot}$),
the companion would have been deeply embedded in the nova envelope,
resulted in strong winds and much shorter outburst (peak plateau timescale of $\sim200$ days).
We note that J0757, with $P_{orb}=119.2$ days and a peak plateau lifetime of $\sim1$ yr (as shown in the lower-left panel in Figure 1),
lies interestingly in the region between the canonical symbiotic novae such as PU Vul,
and novae in close binaries such as V723 Cas.

The absence of emission lines in the spectra of J0757 is peculiar,
but not incompatible with the Hydrogen burning powered outburst picture.
Nova emission lines disappear quite rapidly after the outburst (Downes et al. 2001),
and it is not surprising that we do not see emission lines 60 years after the outburst.
The lack of emission lines could be due to either or both of the following:
The first possible reason is that there is no significant surrounding material.
This may be either due to small amount of mass loss in the system even during the nova outburst,
which is consistent with the recent results by Starrfield et al. (2012);
or due to the fact that ejected shells disappear very fast,
such as observed in V723 Cas (Krautter et al. 2002).
The second possible reason is the lack of ionizing UV photons.
A post-nova WD can cool off fast, for example,
the white dwarf in DQ Her has been found to be below $10\ L_{\odot}$ 40 years after the nova outburst (Stasi\`{n}ska \& Tylenda 1990).
As discussed in sections 2 \& 3, the luminosity of the hot component in J0757
is only a few $L_\odot$,
while the luminosity of the M giant is $\sim250\ L_{\odot}$,
which dominates the spectrum.
Similar cases have been found in a few symbiotic X-ray binaries,
where the hot component is $\sim3$ orders of magnitude fainter than the red giant,
and emission lines are absent in the optical spectra (Masetti et al. 2002, 2007).

In summary, DASCH J0757 is an intriguing system with many unusual properties.
It differs from other known classic or symbiotic novae,
and is more likely a new type of nova-like variable located in the gap region between classic novae and symbiotic novae.
Unlike most other symbiotic stars, it does not contain
a luminous hot component or nebula emission
lines, and its magnitude is stable after the 1940s outburst. Its current photometric
and spectroscopic profile is not different from
a normal red giant binary. It would not be picked
out without the detection of its long outburst in 1940s,
made possible by the long timescales observed with DASCH.
Future observation in the UV and X-ray will help understand its mass transfer and accretion processes.

Given the lack of other observations during the outburst of J0757,
it is hard to reach a definite conclusion whether it was powered by Hydrogen shell burning or accretion.
If it was indeed Hydrogen burning with no significant mass loss,
it could be a promising channel of accumulating mass on the WDs.
Systems similar to DASCH J0757, but with more massive WDs ($>\sim1\ M_{\odot}$), could thus be promising type Ia SNe progenitors.
They are not luminous supersoft X-ray sources in quiescence (see also Di Stefano 2010),
and would not be picked out by the usual searches for symbiotic stars,
but could be detected by the study of their long-term light curves.
Note that J0757 is found from $\sim700$ plates,
while there are nearly half a million plates available to
be scanned and analyzed in DASCH.
With its 100 year light curve coverage,
DASCH has a great potential to find more of such sources,
which will shed light on the understanding of type Ia SNe progenitors,
as well as the study of binary evolution and accretion physics.

\acknowledgments
We thank the anonymous referee for helpful comments.
We are grateful to Alison Doane, Jaime Pepper, Edward Los, Robert J. Simcoe and David Sliski at CfA for their work on DASCH,
and many volunteers who have helped digitize logbooks, clean and scan plates (\emph{http://hea-www.harvard.edu/DASCH/team.php});
and David Latham for his generous help on getting TRES spectra and data reduction.
We thank Sumner Starrfield, Joanna Miko{\l}ajewska, Scott Kenyon, Mariko Kato, Lars Bildsten,
Tony Piro, Sterl Phinney, Ken Shen, Maureen van den Berg, Rosanne Di Stefano, Jing Luan and Perry Berlind for helpful discussions.
S.T. thanks FLWO staff and many visiting astronomers who acquired the TRES and FAST spectra;
Andrea Dupree and Anna Frebel for the help on obtaining and reducing the MIKE spectra;
Jessica Mink and Bill Wyatt for the initial reduction of the TRES and FAST spectra.
This paper uses data products produced by the OIR Telescope Data
Center, supported by the Smithsonian Astrophysical Observatory.
This research has made use of the GSC 2.3.2, 2MASS, SDSS and GALEX catalogs, HEASARC,
NASA/IPAC Infrared Science Archive, and the ASAS database.
This work was supported in part by NSF grants AST0407380 and AST0909073
and now also the \emph{Cornel and Cynthia K. Sarosdy Fund for DASCH}. \\

%% Use the figure environment and \plotone or \plottwo to include
%% figures and captions in your electronic submission.
%% To embed the sample graphics in
%% the file, uncomment the \plotone, \plottwo, and
%% \includegraphics commands
%%
%% If you need a layout that cannot be achieved with \plotone or
%% \plottwo, you can invoke the graphicx package directly with the
%% \includegraphics command or use \plotfiddle. For more information,
%% please see the tutorial on "Using Electronic Art with AASTeX" in the
%% documentation section at the AASTeX Web site,
%% http://www.journals.uchicago.edu/AAS/AASTeX.
%%
%% The examples below also include sample markup for submission of
%% supplemental electronic materials. As always, be sure to check
%% the instructions to authors for the journal you are submitting to
%% for specific submissions guidelines as they vary from
%% journal to journal.

%% This example uses \plotone to include an EPS file scaled to
%% 80% of its natural size with \epsscale. Its caption
%% has been written to indicate that additional figure parts will be
%% available in the electronic journal.

\begin{figure}
\epsscale{.99}
\plotone{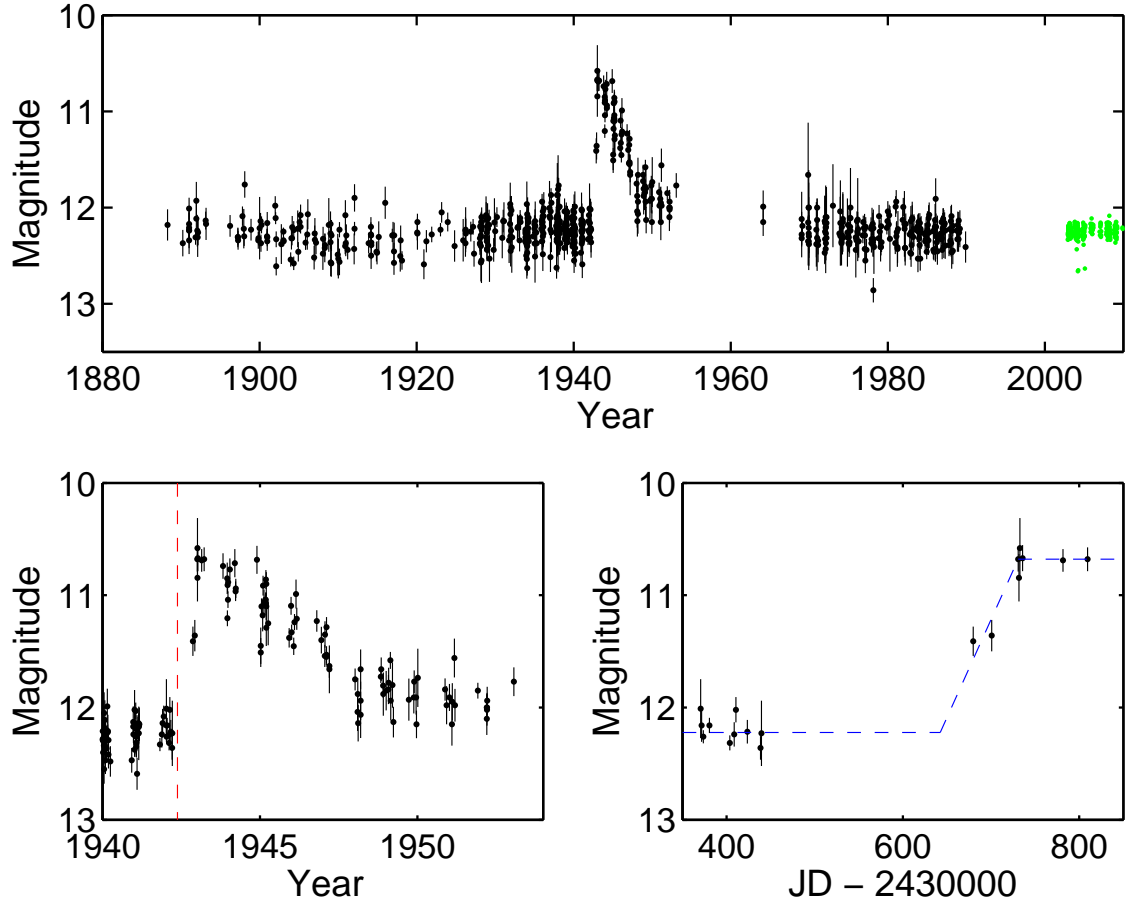}
\caption{Light curve of DASCH J075731.1+201735.
The upper panel shows the whole light curve.
Black dots with error bars are
the light curves from DASCH, and small green dots after Year 2000 are the light curves from ASAS.
Since ASAS data are in the V band, while DASCH magnitudes are in the B band, we added
1.5 mag to the ASAS V mag in the plot to convert it to B, as described in the text.
The lower left panel shows the light curve from 1940 to
1954 for the outburst event; Data before the red dashed line (May 1942) are used as before the outburst.
The lower right panel shows the light curve around 1942 for the rising of the outburst.
The blue dashed line marks the median magnitudes in JD $2430250-2430440$ (12.22 mag; quiescence state)
and JD $2430730-2430850$ (10.68 mag; brightest state),
and a naive linear increase in between; The time span of the linear increase is 88.5 days.
 \label{fig1}}
\end{figure}

\begin{figure}
\epsscale{.9}
\plotone{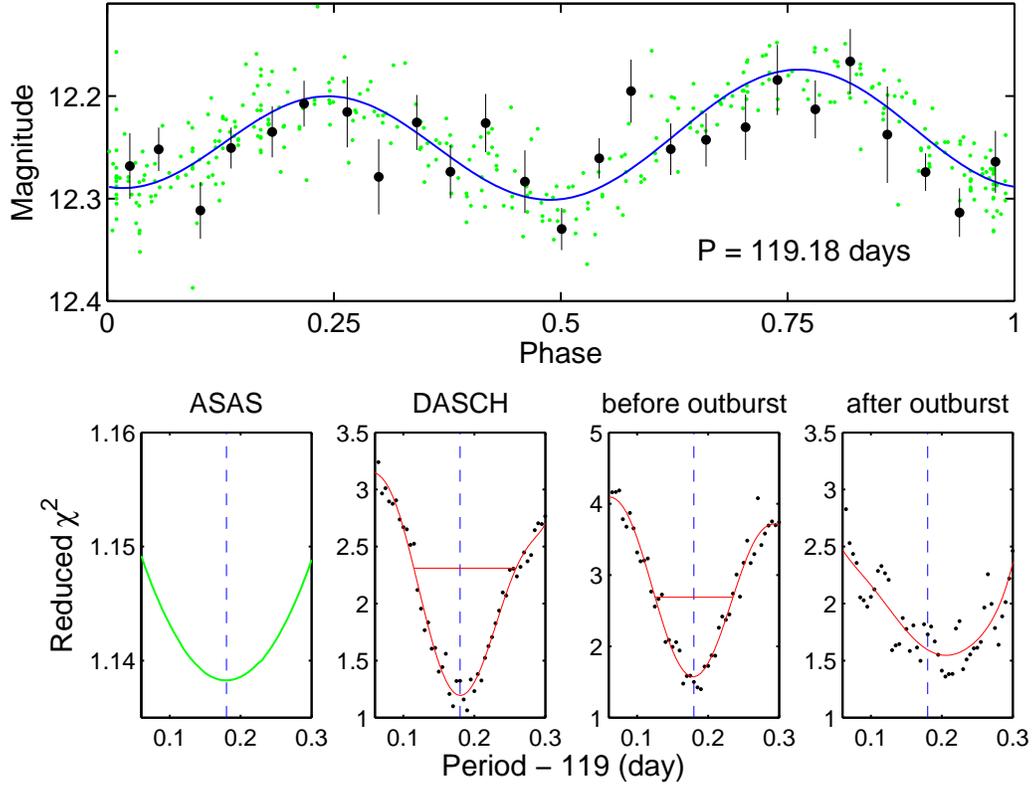}
\caption{Folded light curve and period search of DASCH J0757.
The upper panel shows the foled light curve with $P=119.18$ days and $T_0=2455597.98$.
Black solid dots are DASCH data binned by phase, with error bars represent the standard deviation of the mean.
Small green dots are ASAS data.
Blue solid line is the best fit 3 harmonics Fourier model derived from the ASAS data.
The lower panels show the reduced $\chi^2$ as a function of folding period.
From the left to the right are the reduced $\chi^2$ using ASAS data, DASCH data excluding the outburst,
DASCH data before the outburst (before June 1942, as indicated as red dashed line in the lower left panel in Figure 1),
and DASCH data after the outburst (after 1960). The black dots are the reduced $\chi-$square values, and the red lines are a sixth order polynomial fit.
In the two lower middle panels, i.e. DASCH data excluding the outburst, and DASCH data before the outburst,
$1-\sigma$ confidence levels by assuming $\chi^2$ distribution are marked as red horizontal lines.
The more limited DASCH data after the outburst may suggest a $\sim0.02$ days longer period,
but this is not a significant deviation.
The vertical blue dashed lines mark the best fit period of 119.18 days.\label{fig2}}
\end{figure}

\begin{figure}
\epsscale{.99}
\plotone{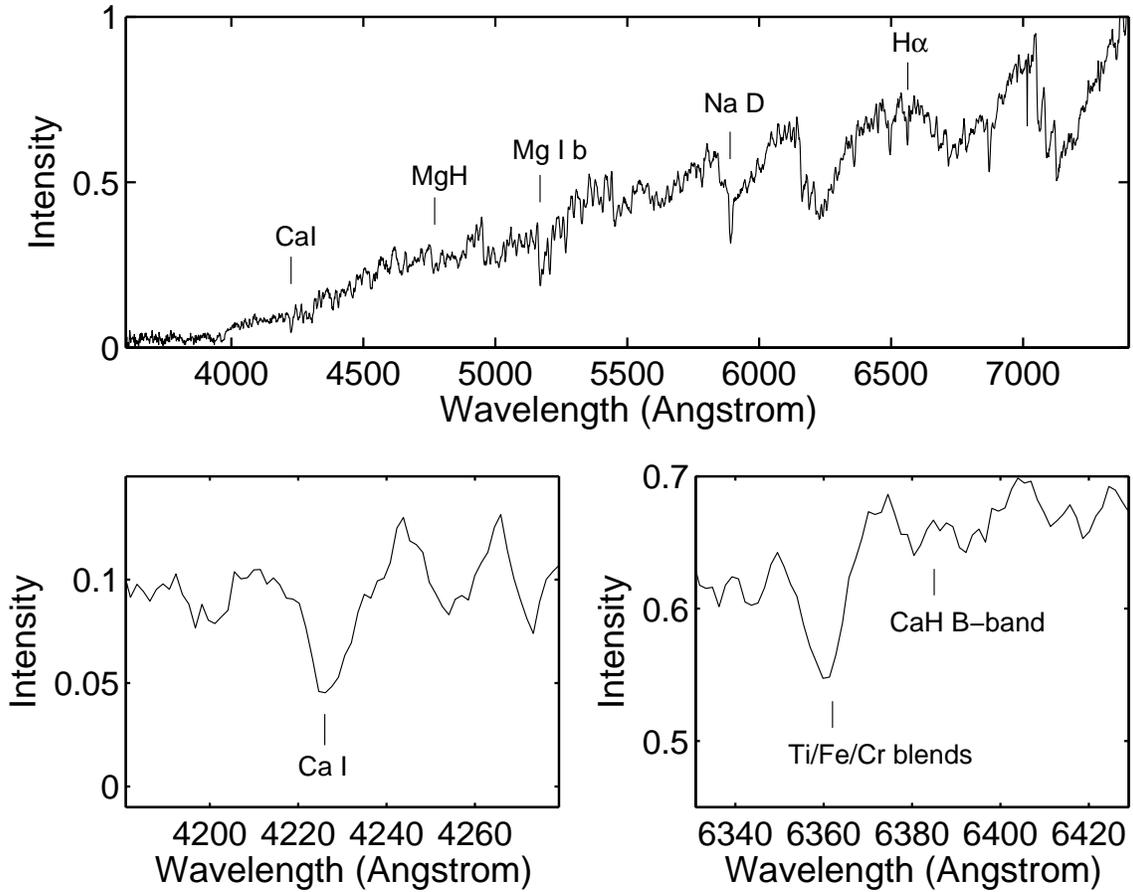}
\caption{FAST 300 grating spectrum of DASCH J0757. Spectral resolution is
about 7 \AA. The upper panel shows the whole spectrum.
The lower panels show two luminosity indicators, Ca I at 4226 \AA,
and the CaH B-band at 6385 \AA\ with the blend near 6362 \AA\ consisting of Ti, Fe, and Cr on the left. \label{fig3}}
\end{figure}

\begin{figure}
\epsscale{.99}
\plotone{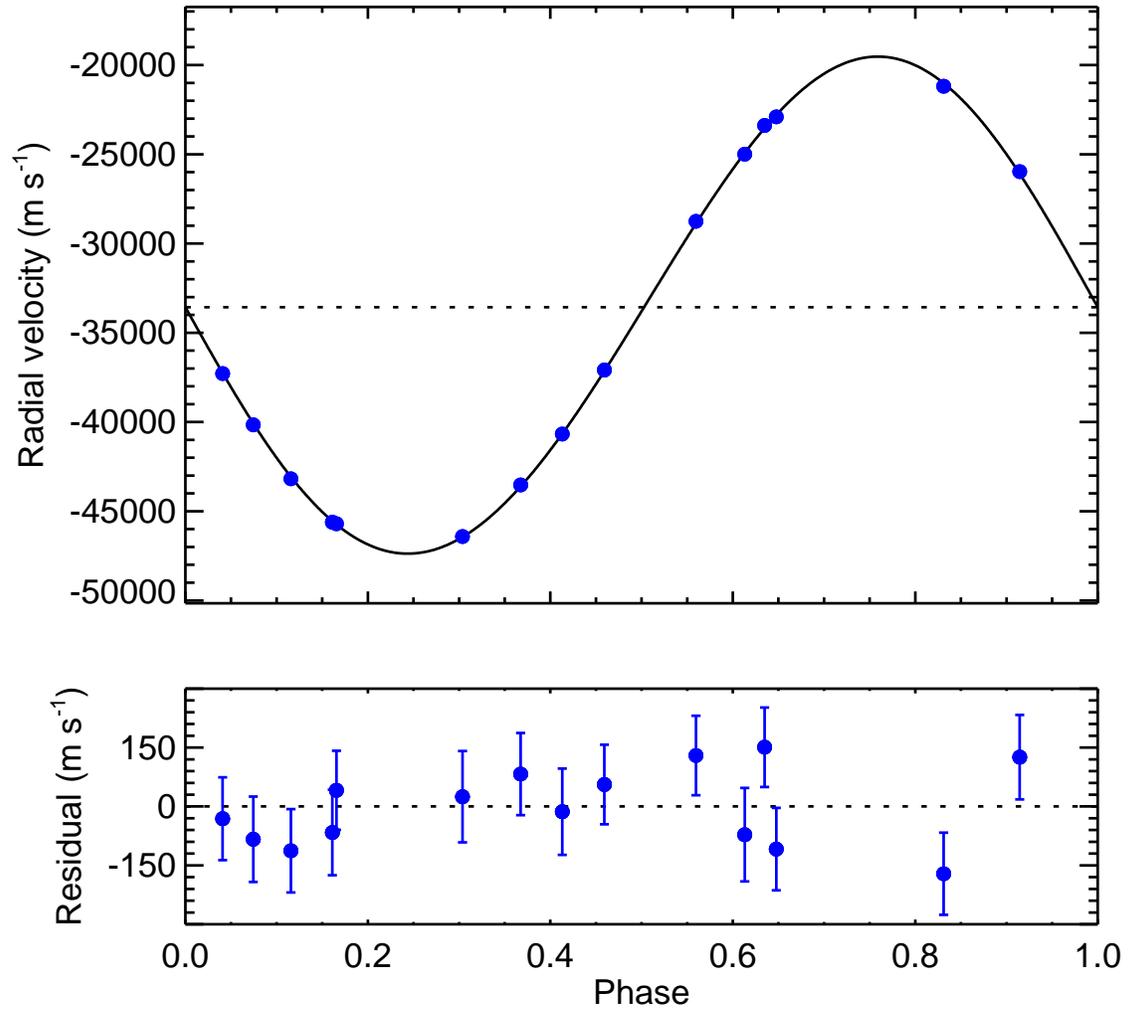}
\caption{Radial velocity of DASCH J0757 as a function of orbital phase, using the ephemeris in Table 1.
Blue solid dots are observation data using the TRES spectrograph at the FLWO 1.5m, and the smooth curve is the best-fitting model.
Residuals from the fit are given in the lower panel. \label{fig4}}
\end{figure}

\begin{figure}
\epsscale{.99}
\plotone{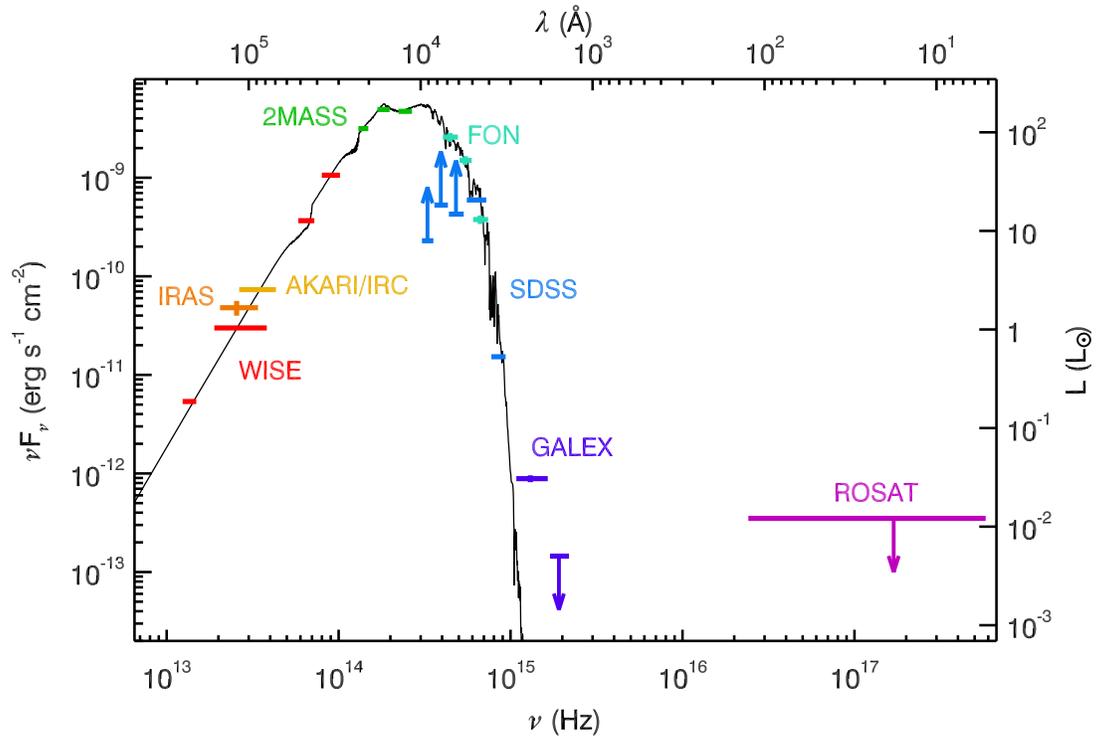}
\caption{SED of DASCH J0757 using archival data. The black line represents a synthetic spectrum of a $T_{eff}=3750$ K,
and log g$ = 1.5$ giant (Lejeune et al. 1997). Note that SDSS riz bands are saturated; It was not detected by GALEX FUV and ROSAT. \label{fig6}}
\end{figure}

\begin{figure}
\begin{center}
\includegraphics[angle=0,width=14cm]{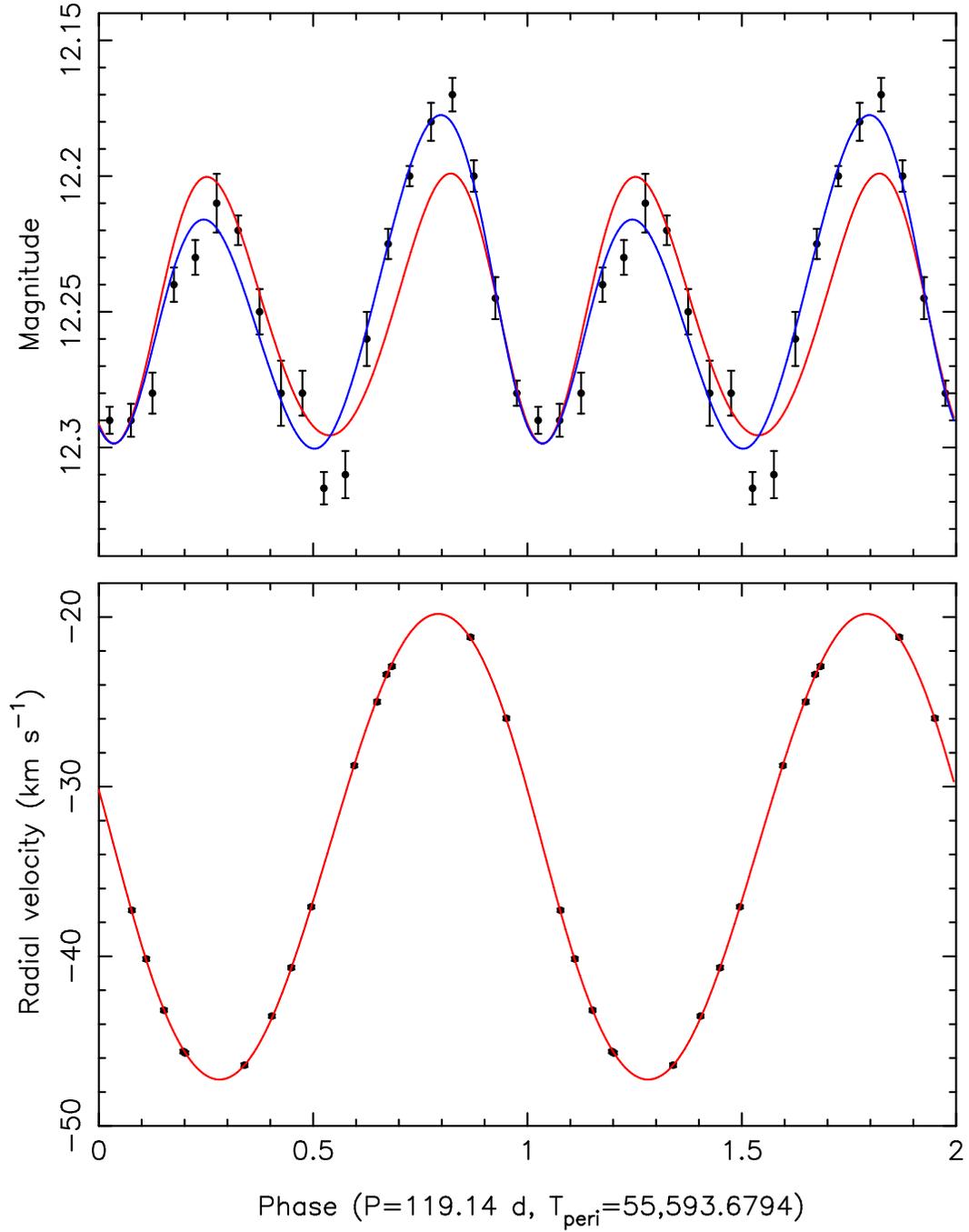}
\caption{Top:  The phased and binned ASAS V-band light curve of J7057
and some representative ellipsoidal models.  The red line is a model
that includes irradiation and non synchronous rotation, and the
blue line is a model that has an accretion disk with a hot spot.
Bottom:  The TRES radial velocity measurements phased on the
same ephemeris (which uses the time of periastron passage as the zero-point)
and the best-fitting model. }
\label{fig5}
\end{center}
\end{figure}

\clearpage

\end{document}